\documentclass{llncs}
\usepackage{makeidx}  
\usepackage{graphicx}
\usepackage{amsmath,amssymb} 
\newcommand{\ket}[1]{ \left |#1\right \rangle}
\makeindex

\begin{document}

\frontmatter          
\setcounter{page}{1}

\pagestyle{headings}  

\title{Measurement-Based Quantum Turing Machines and Questions of Universalities}
\author{Simon Perdrix, Philippe Jorrand}
\institute{Leibniz Laboratory\\ 46, avenue F\'elix Viallet 38000 Grenoble, France\\ \emph{simon.perdrix@imag.fr, philippe.jorrand@imag.fr}}
\maketitle
\begin{abstract}

Quantum measurement is universal for quantum computation (Nielsen \cite{N01}, Leung \cite{L01,L03}, Raussendorf \cite{R00,R03}). This universality allows alternative schemes to the traditional three-step organisation of quantum computation: initial state preparation, unitary transformation, measurement. In order to formalize these other forms of computation, while pointing out the role and the necessity of classical control in measu\-rement-based computation, and for establishing a new upper bound of the minimal resources needed to quantum universality, a formal model is introduced by means of Measurement-based Quantum Turing Machines.

\end{abstract}

\section{Motivations}

\subsection{Preliminaries}

Quantum computation takes advantages of phenomena belonging to quantum mechanics for encoding, processing and communicating information. This allows more efficient computations than those based upon classical mechanics. 

According to the postulates of quantum mechanics, the state of a quantum system is a vector in a Hilbert space. If this system is closed, then its evolution is unitary, hence reversible. Otherwise, the action which consists in opening the system is a measurement. These postulates are reflected by the traditional three-step approach to quantum computation: initial preparation of the quantum system state, application of unitary transformations, measurement of a final state.

We briefly introduce notations which are used for describing quantum states, unitary transformations and quantum measurements, referring to the books by Nielsen and Chuang \cite{NC}, and by Kitaev, Shen and Vyalyi \cite{KSV} for more details. Here, a quantum system is a register of qubits. The state of a $n$-qubit register is a normalized vector in a $2^n$ dimensional Hilbert space $\mathcal{H}_n$. $\mathcal{B}=\{\ket{i}\}_{i\in \{0,1\}^n}$, where $\ket{i}$ is a vector in Dirac's notation, is the computational basis of $\mathcal{H}_n $. Thus a $n$-qubit state $\ket{\phi}$ in the computational basis is: $\ket{\phi}=\Sigma_{i\in\{0,1\}^n}\alpha_i\ket{i}$, $\alpha_i \in \mathbb{C}$. 

A unitary evolution $\mathcal{U}$, which acts on $n$ qubits, is represented by a $2^n\times 2^n$ unitary matrix $U$. $\mathcal{U}$ transforms $\ket{\phi}$ into $U\ket{\phi}$. Clearly, a unitary evolution is deterministic.

A measurement is represented by an observable $O$ which is a hermitian matrix. Considering the spectral decomposition of $O$, $O=\Sigma_mmP_m$, a $O$-measurement transforms, with probability $p_m$, a state $\ket{\phi}$ into the state $\frac{P_m\ket{\phi}}{\sqrt{p_m}}$, where $p_m$ is the scalar product of $\ket{\phi}$ and $P_m\ket{\phi}$. The classical outcome of the measurement is $m$. Thus a measurement is a probabilistic transformation of the state of the register, which returns a classical outcome $m$. The classical outcome gives information on the state of the system after the measurement.

Pauli matrices, $I,X,Y,Z$ are unitary matrices which can be viewed as unitary operators (in this case the notation $\sigma_x,\sigma_y,\sigma_z$ is prefered to $X,Y,Z$). They can also be viewed as observables since they are hermitian. Pauli matrices form a group (up to a global phase), e.g. $X^2=I$ and $X.Y=i.Z$. A $Z$-measurement is also called a measurement in the computational basis.

\begin{center}
$I=\left(\begin{array}{cc}
  1 & 0\\
  0&1\\
\end{array} \right)$, 
$X=\sigma_x=\left(\begin{array}{cc}
  0 & 1\\
  1&0\\
\end{array} \right)$, 
$Y=\sigma_y=\left(\begin{array}{cc}
  0 & -i\\
  i&0\\
\end{array} \right)$, 
$Z=\sigma_z=\left(\begin{array}{cc}
  1 & 0\\
  0&-1\\
\end{array} \right)$
\end{center}
\subsection{Quantum computations are not unitary} 
Unitary transformations characterize the evolution of \emph{closed} systems. For a unitary transformation to be \emph{applied} to a quantum system, an interaction between this quantum system and an experimentalist belonging to the classical world is necessary. This implies that the quantum system is opened, in turn implying that its evolution can no longer be described by postulates on the evolution of \emph{closed} quantum systems.

Even if, from an experimental point of view, it is still reasonable, up to some approximations, to consider that a unitary transformation is nevertheless applied to the quantum system, from a computational point of view, these approximations cannot be accepted, for reasons explained next.

The traditional organization of a quantum computation consists in initialising a quantum system in the zero state (i.e. all qubits are in state $\ket{0}$), then in applying unitary transformations and, at the end of the computation, in measuring the system in the computational basis. In this computational scheme, only the stage of unitary transformations depends on the algorithm (e.g. Shor's algorithm) and on the classical instance of the problem (e.g. a large number to factorize). If the system is considered as being \emph{closed} during this stage of unitary transformations, then the outcome produced by this computation may not depend on the algorithm nor on the problem instance. Information about the algorithm and the classical instance is never \emph{transmitted} to the quantum system because this system is closed, so the outcome of the computation has no interest.   

To sum up, from a strictly abstract computational point of view, closed quantum systems cannot be controlled, because if one tries to control them, they necessarily become open, hence quantum measurements seem to be the unique tool which can be used to perform a quantum computation. Due to the universality of quantum measurements proved by Nielsen \cite{N01}, all unitary transformations may be \emph{simulated} using quantum measurements only. So, even if unitary transformations cannot be used as primitives directly provided by quantum mechanic, all existing algorithms which are based on unitary transformations can be simulated on a measurement-based quantum computer.

\section{Universalities}

\subsection{Definitions}

There exists different abstract models for classical computation (e.g. Turing Machines, Automata) and for quantum computation (e.g. Quantum Circuits, Quantum Turing Machines). We introduce in this section a way of comparing such models with respect to their power. The most powerful models are considered \emph{universal}.

A model is a set of machines. The power of models depends on the \emph{context} of their utilization. Only machines which transform an input $in$ into an output $out$ are considered. Contexts are pairs $In \times Out$ of a set $In$ of inputs and a set $Out$ of outputs. Each machine $m$ is represented by its action, i.e. a relation $m \in \mathcal{P}(In \times Out)$. For all $in \in dom(m)$, where $dom(m)\subset In$ is the domain of definition of $m$, $m(in)\subset Out$ represents the elements of $Out$ which are in relation with $in$ by $m$. In the deterministic case, $m$ is a function, so $m(in)\in Out$.

A machine is an element of $\mathcal{P}(In \times Out)$ (i.e. a subset of $In \times Out$), but not all the elements of $\mathcal{P}(In \times Out)$ are machines, because a machine must be \emph{physically realizable} or, in other words, the relation must be computable.
A model $Mod$ is a set of machines, so $Mod \in \mathcal{P}(\mathcal{P}(In \times Out))$. $\mathcal{R} \subset \mathcal{P}(\mathcal{P}(In \times Out))$ is the set of realizable models, i.e. models comprising physically realizable machines only.

In order to compare the power of two models, a notion of \emph{simulation} is introduced. A machine $m_1$ $S_1$-simulates a machine $m_2$ ($m_2 \prec_{S_1} m_1$) under the context $In \times Out$ iff for all $in \in dom(m_2)$, $m_1(in)=m_2(in)$. Simulation $\prec_{S_1}$ is a quasi-order relation on $\mathcal{P}(In \times Out)$.

In order to obtain a more restrictive simulation $\prec_{S_2}$ (i.e. $\prec_{S_2} \subset \prec_{S_1}$), some additional conditions of complexity may be introduced: a machine $m_1$ $S_2$-simulates a machine $m_2$ under the context $In \times Out$ iff $m_2 \prec_{S_1} m_1$ and $\forall in \in dom(m_2)$, $\tau_{m_1}(in)=O(\tau_{m_2}(in))$, where $\tau_m(in)$ is the execution time of $m$ on $in$. Moreover in order to obtain a less restrictive simulation $\prec_{S_3}$ (i.e. $\prec_{S_1} \subset \prec_{S_3}$), only some conditions of correction may be required: a machine $m_1$ $S_3$-simulates a machine $m_2$ under the context $In \times Out$ iff for all $in \in dom(m_2)$, $m_1(in)\subset m_2(in)$. 

In general, any quasi-order relation $\prec_S$ on $\mathcal{P}(In \times Out)$ is a simulation. This relation $\prec_S$ is extended to models: a model $Mod_1$ $S$-simulates a model $Mod_2$ ($Mod_2 \prec_{S} Mod_1$) under the context $In \times Out$ iff $\forall m_2 \in Mod_2, \exists m_1 \in Mod_1$ such that $m_2 \prec_S m_1$.

For a given context, a model $Mod_1$ is $S$-universal, iff $\forall Mod \in \mathcal{R}, Mod\prec_S Mod_1$.

For a given context, a given model $Mod$ and a given simulation $\prec_S$, $m_1\in Mod$ is a $Mod$-universal machine iff $\forall m \in Mod, \exists in_{m}\in In$ such that $m \prec_S m_1[in_m]$, where $m_1[in_m]$ is $m_1$ with $in_m$ always inserted in the input.

\subsection{Classical Universality}

The context of classical universality is a context of language recognition, i.e. for a finite vocabulary $V$, $In=V^*$ and $Out=\{true, false\}$. Note that the context $In=Out=V^*$ may also be chosen. A machine $m_1$ classically simulates $m_2$ ($m_2\prec_{Class} m_1$), iff for all $in \in dom(m_2), m_1(in)=m_2(in)$.

If $Mod_{TM}$ is the model of classical Turing Machines, the Church-Turing thesis is nothing but: $\forall Mod \in \mathcal{R}, Mod \prec_{Class} Mod_{TM}$. Thus a model $Mod$ is classically universal iff $Mod_{TM}\prec_{Class} Mod$.

Moreover, there exists a machine $m_{univ}\in Mod_{TM}$ which is universal, i.e. $\forall m \in Mod_{TM}, \exists in_{m} \in V^*$ such that $\forall in \in V^*, m \prec_{Class} m_{univ}[in_m]$

\subsection{Quantum Universality}

The context of quantum universality is a context of quantum states transformation, i.e. $In=\mathcal{H}$ and $Out=\mathcal{H}$, where $\mathcal{H}$ is the Hilbert space of quantum states. Note that the context of density matrices transformation may also be chosen. A machine $m_1$ quantum simulates $m_2$ ($m_2\prec_{Quant} m_1$), iff for all $in \in dom(m_2), m_1(in)=m_2(in)$.
\section{Measurement-based Quantum Turing Machines}


 \subsection{Definitions}
 
 A \emph{Measurement-based Quantum Machine} ($MQTM$) is composed of:
 \begin{itemize}
 \item One or several tapes (finite or infinite) of qubits,
 \item One or several measurement heads per tape,
 \end{itemize}
 
 A transition $\delta$ of a $MQTM$ is a function from $\mathcal{S} \times \mathcal{V}$ to $\mathcal{S}\times \mathcal{O} \times \mathcal{D}$, where
 \begin{itemize}
 \item $\mathcal{S}$ is a finite set of states (in the sense of classical automata theorety),
 \item $\mathcal{O}$ is a set of $k$-qubit observables, where $k$ is the number of measurement heads of the machine,
 \item $\mathcal{V}$ is the set of all possible classical outcomes of observables in $\mathcal{O}$,
 \item $\mathcal{D}$ is the set of possible moves of the heads. In the general case, $\mathcal{D}=\mathbb{Z}^k$, where ${\mathbb{Z}}$ is the set of relative integers.  
 \end{itemize}
 
 A configuration is a couple $(s,v)\in \mathcal{S}\times \mathcal{V}$. Given a configuration $(s,v)$, where $v$ is the outcome of the last measurement, the next configuration $(s',v')$ is given by the transition function $\delta$, where $s'\in \mathcal{S}$ and $v'$ is the classical outcome of a measurement performed according to the observable $O\in \mathcal{O}$ and after the measurement heads have been moved according to $d\in \mathcal{D}$.

 A computation on a $MQTM$ $m$ operates as follows:
 \begin{itemize}
 \item The input of the computation is placed on a specified tape, all others qubits are in an unknown state, which is not entangled with input qubits.
 \item A specified measurement head points on the first qubit of the input.
 \item $m$ is placed in an initial configuration $(s_0,v_0)$: $s_0\in \mathcal{S}$ is the initial state and $v_0$ plays the role of the last measurement outcome. 
 \item Then, transitions are successively applied, transforming the state of the tapes.
 \item Computation is terminated when no transition can be applied. At that time, a specified head points on the quantum output of the computation. 
 \end{itemize}

One may note that transitions permit a formal description of communications between the classical and quantum worlds. $M_{MQTM}$ is the model comprising all Measurement-based Quantum Turing Machines operating according to this principle.
  
\subsection{Universality}

In the model of quantum computation by measurement improved by Leung \cite{L01}, the set of two-qubit observables $\{ X\otimes X, Z\otimes Z, X\otimes Z,Z\otimes X, X\otimes I,Z\otimes I, \frac{1}{\sqrt{2}}(X\otimes X+Y\otimes X)  \}$ is proved to be quantum universal. This leads to the following Lemma, which is proved in the appendix:

\begin{lemma}
The model $M_A\subset M_{MQTM}$ of $MQTM$ composed of one infinite tape, two measurement heads, $\mathcal{D}_A=\mathbb{Z}^2$ and $\mathcal{O}_A=\{ X\otimes X, Z\otimes Z, X\otimes Z,Z\otimes X, X\otimes I,Z\otimes I, \frac{1}{\sqrt{2}}(X\otimes X+Y\otimes X),\frac{1}{\sqrt{2}}(X\otimes X+X\otimes Y)    \}$, is quantum universal.
\end{lemma}
\begin{center}
\includegraphics[width=0.4\textwidth]{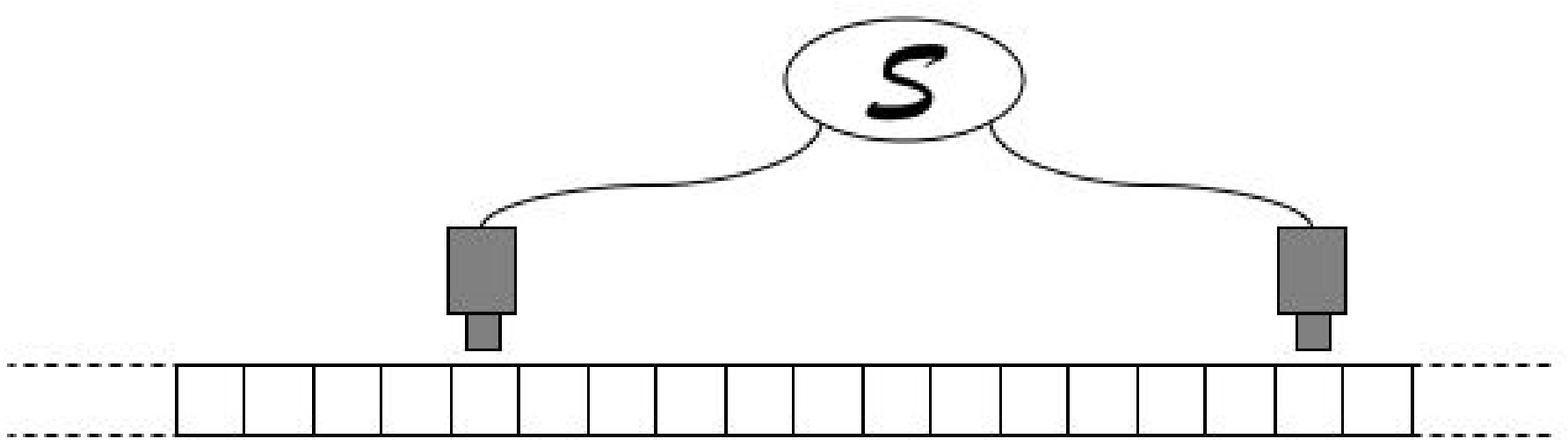}

\emph{\small Figure 1. Machines in model $M_A$}
\end{center}
 \begin{lemma}
 For any set $\mathcal{O}_B$ of $1$-qubit observables and for any $\mathcal{D}_B \subset \mathbb{Z}$ , the model $M_B\subset M_{MQTM}$ of $MQTM$ composed of one infinite tape, one measurement head, $\mathcal{O}_B$ and $ \mathcal{D}_B$, is \emph{not} quantum universal.  
\end{lemma}
 
 \begin{proof}
By counter-example. Entanglement can not be created using only one-qubit measurements, i.e. for a given register of two qubits in the state $\ket{00}$ (which is a separable state) and for any sequence $s$ of one-qubit measurements, the state of the register after the application of $s$ is separable. Thus the unitary transformation $U=(H\otimes I).CNot$, which transforms the separable state $\ket{00}$ into the entangled state $(\ket{00}+\ket{11})/\sqrt{2}$ can not be simulated by a machine of $M_B$, whereas there exists $m_A\in M_A$ which simulates $U$. So $M_B$ is not quantum universal.
 $\hfill \Box$
 \end{proof}
 
 \begin{center}
\includegraphics[width=0.4\textwidth]{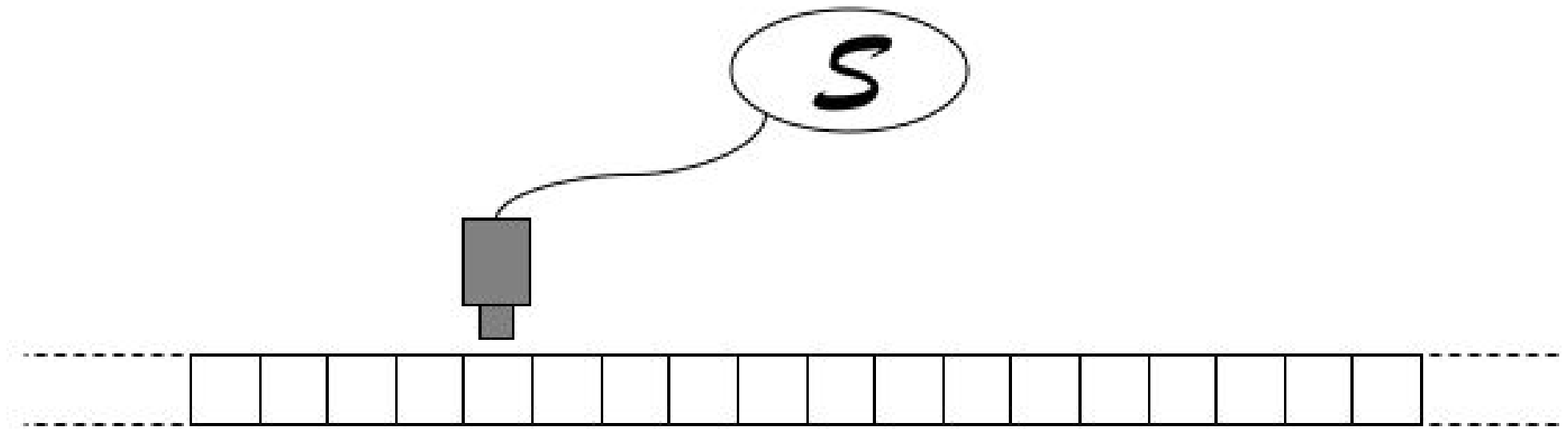}

\emph{\small Figure 2. Machines in models $M_B$ and $M_C$}
\end{center}

\begin{lemma}
The model $M_C\subset M_{MQTM}$ of $MQTM$ composed of one infinite tape, one measurement head, $\mathcal{D}_C=\{-1,0,+1\}$ and $\mathcal{O}_C=\{ X, Z\}$, is classically universal, i.e. $M_{TM} \prec_{Class} M_C$.
\end{lemma}

 \begin{proof} We prove that $M_C$ classically simulates the model $M_{TM}$ of classical Turing Machines composed of an infinite tape of bits and a read-write head, i.e for any classical Turing Machine $m\in M_{TM}$, there exists a machine $m_C\in M_C$, such that $m\prec_{Class} m_C$. For a given $m\in M_{TM}$, a machine $m_C \in M_C$ is considered, such that: the tape of qubits plays the role of the tape of bits; the classical value $0$ (resp $1$) is represented by the quantum state $\ket{0}$ (resp $\ket{1}$). In order to simulate classical reading, a $Z$-measurement is performed: if the state of the qubit is $\ket{0}$ the classical outcome of the measurement is $0$ with probability one, same for $\ket{1}$ with the classical outcome $1$. Thus the measurement head of $m_C$ plays the role of the reading head of $m$. In order to simulate writing, for instance of the value $0$ on a bit, an $X$-measurement followed by a $Z$-measurement are performed. After the $X$-measurement, the state of the qubit is $(\ket{0}+\ket{1})/\sqrt{2}$ or $(\ket{0}-\ket{1})/\sqrt{2}$, and, after the $Z$-measurement, is $\ket{0}$ with probability $1/2$, and $\ket{1}$ with probability $1/2$. If the state is $\ket{1}$, the process ($X$-measurement followed by $Z$-measurement) is repeated, until it becomes $\ket{0}$.
 $\hfill \Box$
 \end{proof}
 
 $M_C$ is classically universal (\emph{Lemma 3}), but $M_C$ is not quantum universal (\emph{Lemma 2}), so this model points out a gap between classical computation and quantum computation. From a decidability point of view, quantum and classical computation are equivalent \cite{D85}, which seemed to imply that the only differece concerns complexity issues. However, in our definitions for quantum and classical universalities, there are no restrictions on complexity, therefore one may wonder why we find this gap between quantum and classical universalities. The key is that contexts of utilization differ: for classical universality, machines act on classical inputs, for quantum universality, machines act on quantum inputs.

\begin{lemma}
The model $M_D\subset M_{MQTM}$ of $MQTM$ composed of two infinite tapes, one measurement head per tape, $\mathcal{D}_D=\mathbb{Z}^2$ and $\mathcal{O}_D=\{ X\otimes X, Z\otimes Z, X\otimes Z,Z\otimes X, X\otimes I,Z\otimes I, I\otimes X,I\otimes Z, \frac{1}{\sqrt{2}}(X\otimes X+X\otimes Y),\frac{1}{\sqrt{2}}(X\otimes X+Y\otimes X)   \}$, is quantum universal, i.e. $M_A \prec_{Quant} M_D$.
\end{lemma}
\begin{center}
\includegraphics[width=0.4\textwidth]{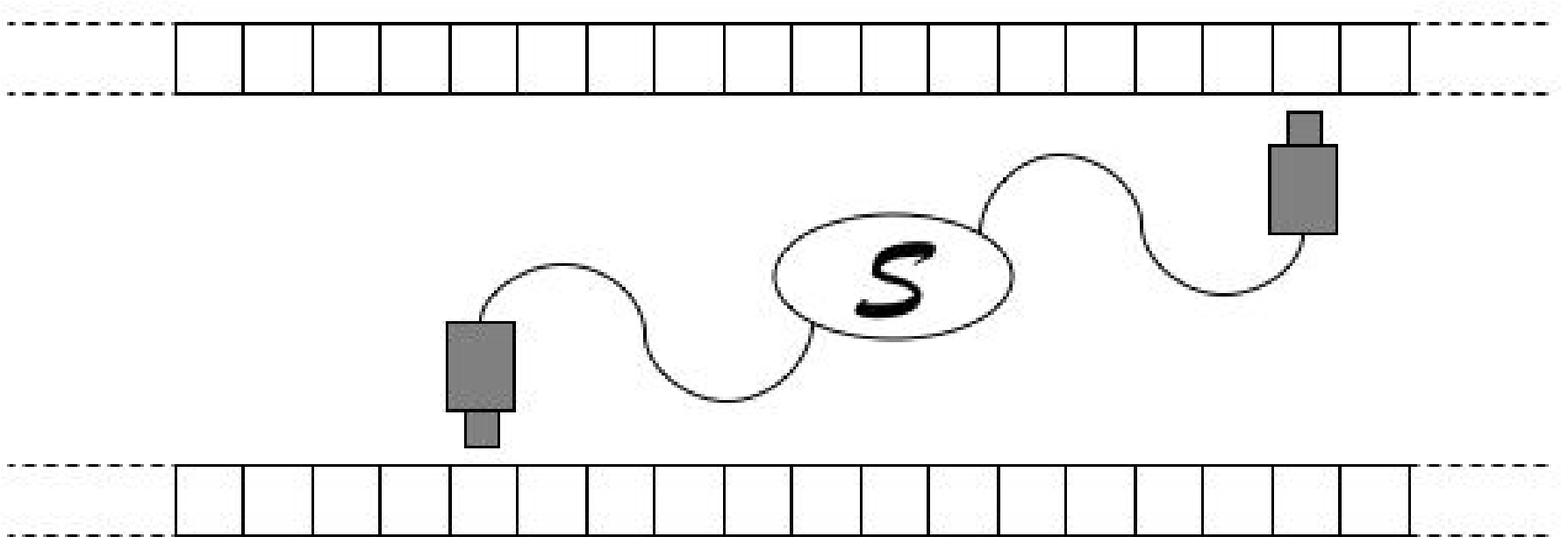}

\emph{\small Figure 3. Machines in model $M_D$}

\end{center}
\begin{proof}
 We prove that $M_D$ quantum simulates $M_A$, i.e for any machine $m_A\in M_A$, there exists a machine $m_D\in M_D$, such that $m_A\prec_{Quant} m_D$.
Qubits of $m_A$ are indexed from $-\infty$ to $+\infty$.
The machine $m_D$ has two tapes: its upper tape and its lower tape (see fig.3). A subset of the qubits of the upper tape of $m_D$ are indexed from $-\infty$ to $+\infty$, using odd numbers only, while a subset of the qubits of the lower tape are numbered with even numbers only, such that there remains an infinite number of non-indexed qubits on each tape of $m_D$. These non-indexed qubits will be available as \emph{auxiliary qubits}.

An execution on $m_A$ is entirely described by a sequence of measurements. We show that each $O$-measurement in this sequence may be simulated by $m_D$. A two-qubit $O$-measurement of $m_A$ acts on qubits of $m_A$ indexed by $i$ and $j$.
\begin{itemize}
\item If $i$ and $j$ have a different parity, a $O$-measurement on qubits $i$ and $j$ is allowed on $m_D$, because $i$ and $j$ are not on the same tape and $O \in \mathcal{O}_D$.
\item Otherwise, assume $i$ and $j$ are both even (so $i$ and $j$ are on the lower tape of $m_D$, see fig. 4). If the state of $j$ is teleported \cite{B93} from the lower tape to an auxiliary qubit $a$ of the upper tape, then $O$ may be applied on $i$ and $a$. A second teleportation from $a$ to $j$ will then terminate the simulation of a $O$-measurement on $i$ and $j$. Thus, the quantum universality of $M_D$ is reduced to the ability to teleport the state of a qubit from one tape of $m_D$ to the other.
\begin{center}
\includegraphics[width=0.4\textwidth]{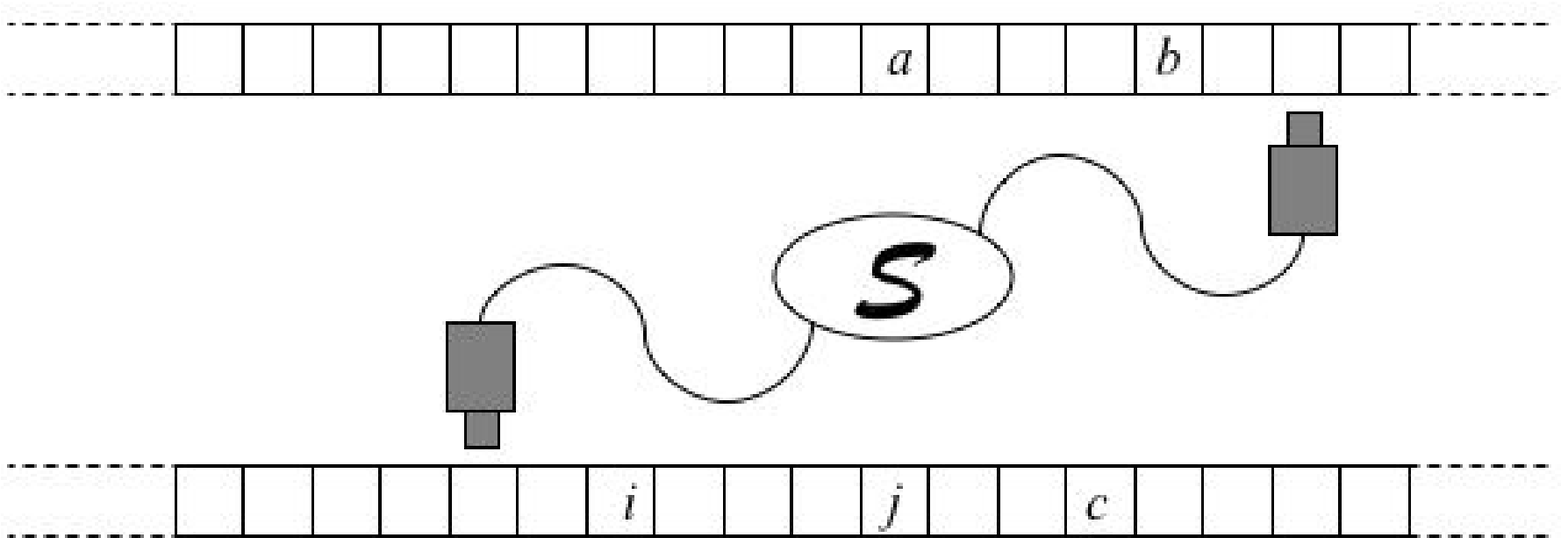}

\emph{\small Figure 4. Simulation of $m_A$ by $m_D$}

\end{center}
Considering a qubit $j$ of the lower tape, two auxiliary qubits $a$ and $b$ of the upper tape, and an auxiliary qubit $c$ of the lower tape, teleporting $j$ to $a$ consists in assigning a Bell state to $a$ and $b$, then performing a Bell measurement on $j$ and $b$. A Bell measurement may be decomposed into a $Z\otimes Z$-measurement followed by a $X\otimes X$-measurement. Applying a Bell measurement on $a$ and $b$ assigns a Bell state to these two qubits, but $a$ and $b$ are both on the upper tape, so a measurement on these two qubits cannot be performed, that is why an auxiliary qubit $c$ is needed. Using $c$, the sequence of measurements $\{I \otimes Z^{(a)},I \otimes Z^{(b)},Z^{(c)}\otimes I, X^{(c)} \otimes X^{(a)}, X^{(c)} \otimes X^{(b)},Z^{(c)} \otimes I\}$, assigns a Bell state to qubits $a$ and $b$. 

The state of the 3-qubit register $a,b,c$ after the first three measurements in this sequence is $\ket{\psi}=(\sigma_x^{\frac{1-i}{2}}\otimes \sigma_x^{\frac{1-j}{2}}\otimes \sigma_x^{\frac{1-k}{2}})\ket{000}$, where $i,j,k\in\{-1,1\}$ are the respective classical outcomes of these measurements. Then the evolution of $\ket{\psi}$ with the remaining measurements in the sequence is:

$\ket{\psi_1}=(\sigma_z^{\frac{1-l}{2}}\sigma_x^{\frac{1-i}{2}}\otimes \sigma_x^{\frac{1-j}{2}}\otimes  \sigma_x^{\frac{1-k}{2}})[\frac{1}{\sqrt{2}}(\ket{000}+\ket{101})]$

$\ket{\psi_2}=(\sigma_z^{\frac{1-l}{2}}\sigma_x^{\frac{1-i}{2}}\otimes \sigma_z^{\frac{1-m}{2}}\sigma_x^{\frac{1-j}{2}}\otimes \sigma_x^{\frac{1-k}{2}})[\frac{1}{2}(\ket{000}+\ket{011}+\ket{101}+\ket{110})]$

$\ket{\psi_3}=( \sigma_x^{\frac{1-k}{2}}\sigma_z^{\frac{1-l}{2}}\sigma_x^{\frac{1-i}{2}}\otimes  \sigma_x^{\frac{1-n}{2}}\sigma_z^{\frac{1-m}{2}}\sigma_x^{\frac{1-j}{2}}\otimes  \sigma_x^{\frac{1-n}{2}}\sigma_x^{\frac{1-k}{2}})[\frac{1}{\sqrt{2}}(\ket{00}+\ket{11})\otimes\ket{0}]$

where $l,m,n \in \{-1,1\}$ are the respective classical outcomes of the last three measurements. At the end, the state of $a$ and $b$ is a Bell state.

Because of the probabilistic aspect of quantum measurement, teleportation succeeds with probability $1/4$. If it does not succeed, a process of correction which consists in teleporting the state of $a$ to another auxiliary qubit, is repeated until a satisfactory state is produced.  $\hfill \Box$
\end{itemize}

\end{proof}

\begin{lemma}
The model $M_E\subset M_{MQTM}$ of $MQTM$ composed of a two-qubit tape, an infinite tape, one measurement head per tape, $\mathcal{D}_E=\{-1,0,1\}\times \mathbb{Z}$ and $\mathcal{O}_E=\{ X\otimes X, Z\otimes Z, X\otimes Z,Z\otimes X, X\otimes I,Z\otimes I, I\otimes X,I\otimes Z, \frac{1}{\sqrt{2}}(X\otimes X+X\otimes Y),\frac{1}{\sqrt{2}}(X\otimes X+Y\otimes X)   \}$, is quantum universal, i.e. $M_A \prec_{Quant} M_E$.
\end{lemma}
\begin{center}
\includegraphics[width=0.4\textwidth]{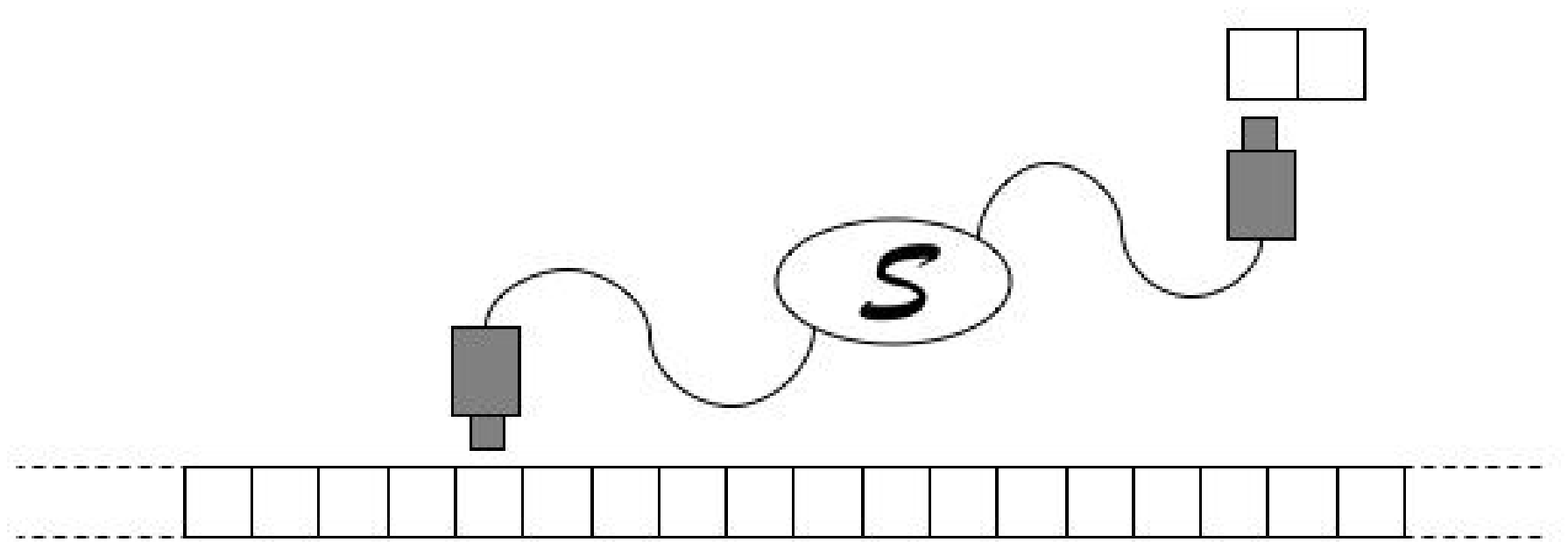}

\emph{\small Figure 5. Machines in model $M_E$}

\end{center}

\begin{proof} 
We prove that $M_E$ quantum simulates $M_A$, i.e for any machine $m_A\in M_A$, there exists a machine $m_E\in M_E$, such that $m_A\prec_{Quant} m_E$.
Qubits of $m_A$ are indexed from $-\infty$ to $+\infty$. A subset of the qubits of the infinite tape of $m_E$ are indexed from $-\infty$ to $+\infty$ such that there remain an infinite number of non-indexed qubits on the infinite tape of $m_E$, which will be available as \emph{auxiliary qubits}.

An execution on $m_A$ is entirely described by a sequence of measurements. We show that each $O$-measurement in this sequence may be simulated by $m_E$. A two-qubit $O$-measurement acts on qubits of $m_A$ indexed by $i$ and $j$. Qubits of the finite tape of $m_E$ are indexed by $a$ and $b$. To simulate a $O$-measurement on $i$ and $j$, the state of $i$ is teleported to $a$, then $O$ is performed on $a$ and $j$, then the state of $a$ is teleported to $i$. Thus the quantum universality of $M_E$ is reduced to the ability to teleport a state from the infinite tape to the finite tape and vice-versa. In the proof of \emph{Lemma 4}, we have seen that the teleportation of a state from one tape to the other needs two qubits on each tape, including the teleported qubit. Thus $M_E$ quantum simulates $M_A$.$\hfill \Box$
\end{proof}

\begin{theorem}
The model $M_F\subset M_{MQTM}$ of $MQTM$ composed of a one-qubit tape, an infinite tape, one measurement head per tape, $\mathcal{D}_F=\{0\}\times \mathbb{Z}$ and $\mathcal{O}_F=\{ X\otimes X, Z\otimes Z, X\otimes Z,X\otimes Z, X\otimes I,Z\otimes I, I\otimes X,I\otimes Z, \frac{1}{\sqrt{2}}(X\otimes X+X\otimes Y)\}$, is quantum universal.
\end{theorem}
\begin{center}
\includegraphics[width=0.4\textwidth]{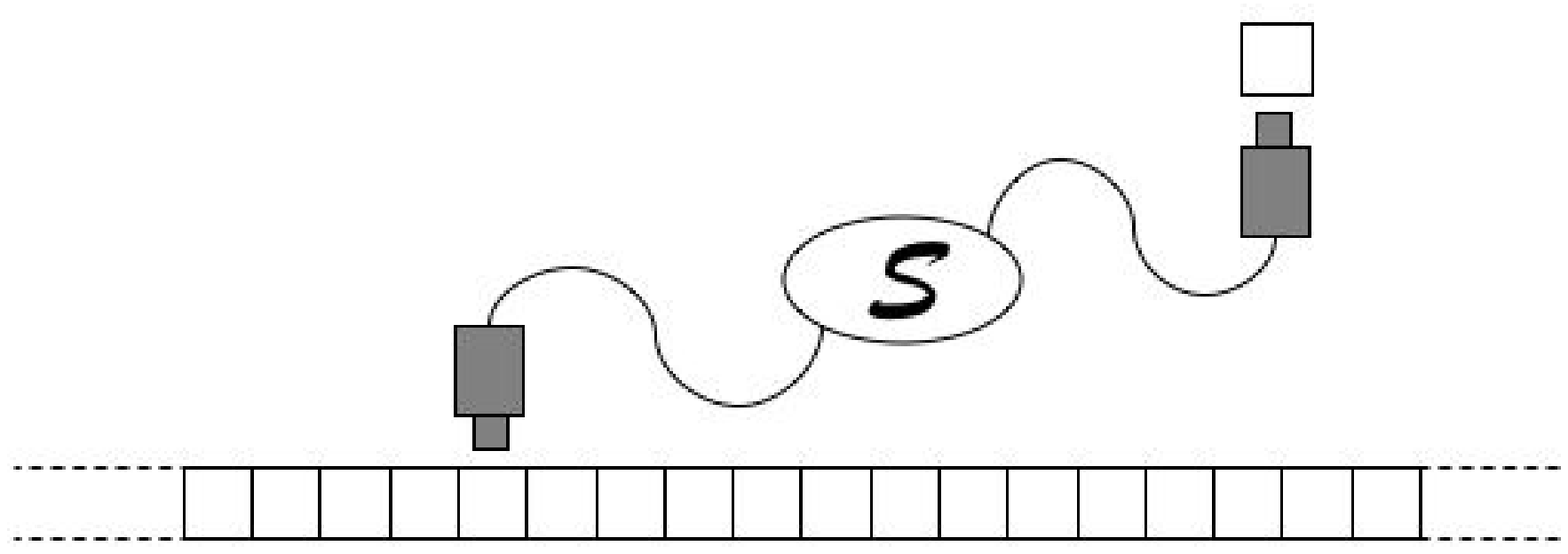}

\emph{\small Figure 6. Machines in the model $M_F$, with the minimal resources for quantum universality}

\end{center}
\begin{proof} Like in \emph{Lemma 5}, the proof is based on the ability to transfer the state of a qubit $j$ of the infinite tape to the qubit $a$ of the other tape and vice-versa. State transfer from $j$ to $a$ consists in $Z$-measuring $a$, then $X\otimes X$-measuring $a$ and $j$, then $Z$-measuring $j$. For a given state $\ket{\psi}=(\alpha\ket{0}+\beta \ket{1})\otimes (\gamma \ket{0} +\delta \ket{1})$ of the 2-qubit register $j,a$, the evolution of $\ket{\psi}$ along the sequence of measurements is:

$\ket{\psi_1}=(I\otimes \sigma_x^{\frac{1-i}{2}})[(\alpha \ket{0}+\beta \ket{1})\otimes \ket{0}]$

$\ket{\psi_2}=(I\otimes \sigma_z^{\frac{1-j}{2}}\sigma_x^{\frac{1-i}{2}})[\alpha \ket{00}+\beta \ket{01}+\beta \ket{10}+\alpha \ket{11}]$

$\ket{\psi_3}=(\sigma_x^{\frac{1-k}{2}}\otimes \sigma_x^{\frac{1-k}{2}}\sigma_z^{\frac{1-j}{2}}\sigma_x^{\frac{1-i}{2}})[\ket{0}\otimes (\alpha \ket{0}+\beta \ket{1})]$

where $i,j,k\in\{-1,1\}$ are the respective classical outcomes of three successive measurements. In order to transfer a state from $a$ to a qubit of the infinite tape the same scheme is applied. The state transfer succeeds up to a Pauli operator. If this Pauli operator is not $I$ then a correcting process is performed, consisting in transferring the state again. $\hfill{\Box}$
\end{proof}

\subsection{Execution tree}

For a given $MQTM$ $m$, a configuration is a couple $(s,v)\in\mathcal{S}\times \mathcal{V}$. Although $\delta$ is a function, an execution on $m$ is \emph{a priori} non deterministic because of quantum measurement. An execution tree $\mathcal{T}_m$ of $m$ can be built, where nodes represent configurations and arcs represent the transitions applied. If $m$ is a $k$-head machine, then each node has at most $2^k$ sons, because each measurement has at most $2^k$ different classical outcomes, so $\mathcal{T}_m$ is a $2^k$-ary tree. 

An execution is a path from the root, which is the initial configuration of $m$, to a leaf, which is a configuration from which no transition may be performed. But an execution may also be a semi-infinite path from the root if the execution never terminates. This would be the case, for instance, when correcting processes fail to succeed infinitely many times.

\subsection{Quantum limitations due to classical control}

In Measurement-based Quantum Turing Machines, transitions permit a formalization of the \emph{communications} which are required for performing and controlling quantum computations. From a computational theoretic point of view, it has been shown in section $1$ that no unitary transformation may actually be applied, hence only quantum measurements are allowed. But one may wonder whether all quantum measurements can be used.

For a given Measurement-based Quantum Turing Machine $m$, it comes from \cite{N97,O98} that if $m$ has an infinite number of measurement heads, then no-computable functions, like the \emph{halting function}, may be performed on it. Therefore, one may conclude that a realizable machine has a finite number $k$ of measurement heads. 

Then, with only $k$-qubit observables, if $\mathcal{V}$ is chosen equal to $\{0\ldots 2^k-1\}$, then $\mathcal{V}$ and $\mathcal{S}$ are finite, so the transition function $\delta$ has a finite domain of definition $\mathcal{S}\times \mathcal{V}$, therefore only a finite part of $\mathcal{O}$ may be used.

But if $\mathcal{V}$ is infinite, for instance $\mathcal{V}=\mathbb{Z}$, choosing $\mathcal{O}=\{O_i, i\in\mathbb{N}\}$ such that for all $i$, the eigenvalues of $O_i$ are $i$ and $-i$, then a transition function like $\forall i\in \mathbb{Z}, \delta (s,i)=(s,O_{\left | i\right |+1},0)$ would explore an infinite number of observables.

Since, for a given $k$-head $MQTM$ $m$, $\mathcal{T}_m$ is $2^k$-ary, a breadth-first walk across $\mathcal{T}_m$ recursively enumerates the nodes of $\mathcal{T}_m$. Moreover, each node $n$ of $\mathcal{T}_m$ may be labeled also with an observable $O_n$ according to $\delta$, i.e. if $n=(s,v)$ then $O_n$ is the observable given by $\delta(s,v)$. Therefore a given Measurement-based Quantum Turing Machine may use only a recursively enumerable set of observables. Hence, the study of $M_{MQTM}$ may be limited to machines on which $\mathcal{O}$, therefore $\mathcal{V}$ are recursively enumerable.


\section{Conclusion}

This paper introduces a unified formalization of the concepts of classical and quantum universalities. We have pointed out, from a strictly abstract computational point of view, that unitary-based quantum computation do not have an adequate and feasible physical realisation. Still from the same point of view, only measurement-based quantum computations can be considered adequate. This led to the introduction of a new abstract model for quantum computations, the model of Measurement-based Quantum Turing Machines ($M_{MQTM}$).

This model allows a rigorous formalization of the necessary interactions between the quantum world and the classical world during a computation. $M_{MQTM}$ has been studied within a general framework of abstract computation models, with a notion of simulation among models allowing to compare them in terms of their universalities.

Two main results have been obtained in this framework. A hierarchy of models contained in $M_{MQTM}$, ranked according to the resources they use, have been proved universal for quantum computations. One of them exhibits a new upper bound for the minimal resources required for quantum universality.

Another subset of $M_{MQTM}$, with more restricted resources, has been proved universal for classical computations, and proved not universal for quantum computations, thus pointing out a gap between the classical and quantum approaches to computing.

\section*{Appendix}
Proof of \emph{Lemma 1}:

For proving that $M_A$ is quantum universal, we prove $M_{Circuit}\prec_{Quant}M_A$, where $M_{Circuit}$ is the model of quantum circuits based on the universal family $\{H,T, CNot\}$ of unitary gates.

A quantum circuit $c$ is a sequence $U_1[A_1],\ldots, U_L[A_L]$, where $U_j\in \{H,T,CNot\}$ and $A_j$ is an $n$-tuple of qubits, where $n\in\{1,2\}$ is the arity of $U_j$: $U_j[A_j]$ means $U_j$ \emph{applied} to the qubits of $A_j$. The number $L$ is called the size of the circuit, we refer the reader to \cite{KSV} for more details on quantum circuits.

The simulation of a circuit $c$ by measurement-based quantum computation (see \cite{L01}), may never terminate. The proof given now focuses on one point: the set $\mathcal{S}$ of states of the machine $m\in M_A$ which simulates $c$ needs only be finite. Technical issues concerning the movement of measurement heads are omitted in this presentation of the proof. With this simplification, a machine $m\in M_A$ is described by an automaton $\mathcal{A}_m$: the states of $\mathcal{A}_m$ are nothing but the states of $m$, and the transitions of $\mathcal{A}_m$, which may be labeled by $v\in \mathcal{V}$, correspond to the transition function $\delta$ of $m$, i.e. there is a transition of $\mathcal{A}_m$, labeled by $v$, from $s_1$ to $s_2$, iff $\exists (O,d)\in \mathcal{O}\times \mathcal{D}$ such that $\delta(s_1,v)=(s_2,O,D)$.

This proof is based on a 3-level decomposition of the problem: \emph{steps of simulation of U}, where a unitary transformation $U$ is simulated up to a Pauli operator (i.e. a step of simulation of $U$ on $\ket{\phi}$ produces $\sigma U \ket{\phi}$, where $\sigma$ is a Pauli operator); \emph{full simulation of U}, which combines steps of simulation; and \emph{simulation of $c$} which combines full simulations of unitary transformations.

The size $L$ of the circuit is finite, implying that the simulation of $c$ with a $MQTM$ $m$ may be decomposed into $L$ full simulations, one for each unitary transformation $U_j$.

\begin{center}
\includegraphics[width=0.6\textwidth]{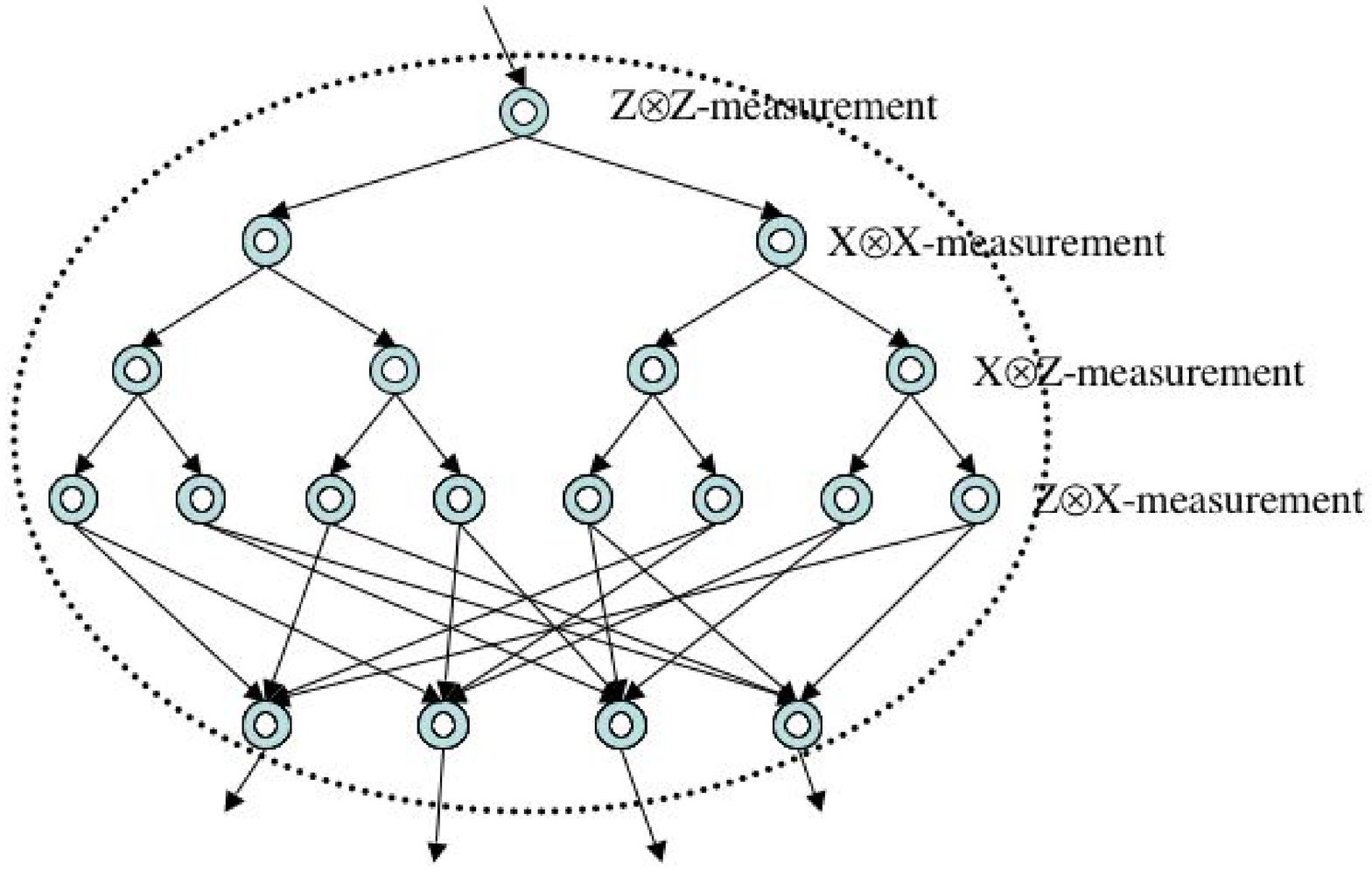}

\emph{\small Figure 7: Detailed step of simulation of $H$.}

\includegraphics[width=0.15\textwidth]{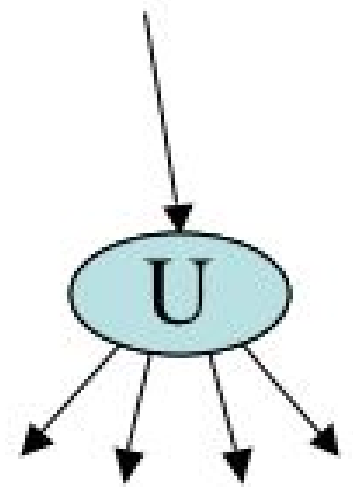}

\emph{\small Figure 8: General black-box representation of a step of simulation of $U$.}
\end{center}

According to the principles of measurement-based quantum computation \cite{L01}, a step of simulation of $U_j$ on $A_j$ consists in preparing an ancilla state which depends on $U_j$, then in performing Bell measurements between some qubits of $A_j$ and some qubits of the ancilla.
For instance, if $U_j=H$, then from a $Z\otimes Z$-measurement followed by a $X\otimes X$-measurement, the ancilla state is obtained. Then a $X\otimes Z$-measurement followed by a $Z \otimes X$-measurement is performed. Each of these four measurements has two possible outcomes, \emph{a priori} giving rise to 16 final states for the automaton which controls this sequence of measurements.
However, these 16 final states may be grouped into four states only, because the simulation of $H$ succeeds up to one of the four Pauli operators.
Thus, in general, the automaton for a step of simulation of a unitary transformation $U$ has one entry state and four final states corresponding to the four Pauli operators that should be applied next. For all $U \in \{H,T,CNot,\sigma_x,\sigma_y,\sigma_z\}$, it is obvious that a step of simulation of $U$ needs a finite number of states in the corresponding automaton (the case of $H$ is shown in fig. $7$). This automaton can be abstracted into a single node (fig $8$) with one incoming transition and four outgoing transitions, for use at the next higher level which is the full simulation of $U$. 

For a given step of simulation of $U$, the full simulation of $U$ is given by an automaton where each state encapsulates a step of simulation (fig.$9$). This automaton is interpreted as follow: $U$ is simulated on a quantum state $\ket{\phi}$, so $\sigma U\ket{\phi}$ is obtained where $\sigma$ is a Pauli operator. If $\sigma=I$ then the simulation is terminated, otherwise $\sigma$ is simulated. From this step of simulation, the state $\sigma'\sigma \sigma U\ket{\phi}=\sigma' U\ket{\phi}$ is obtained. If $\sigma'=I$ the simulation is terminated, otherwise $\sigma'$ is simulated, and so on.
\begin{center}
\includegraphics[width=0.4\textwidth]{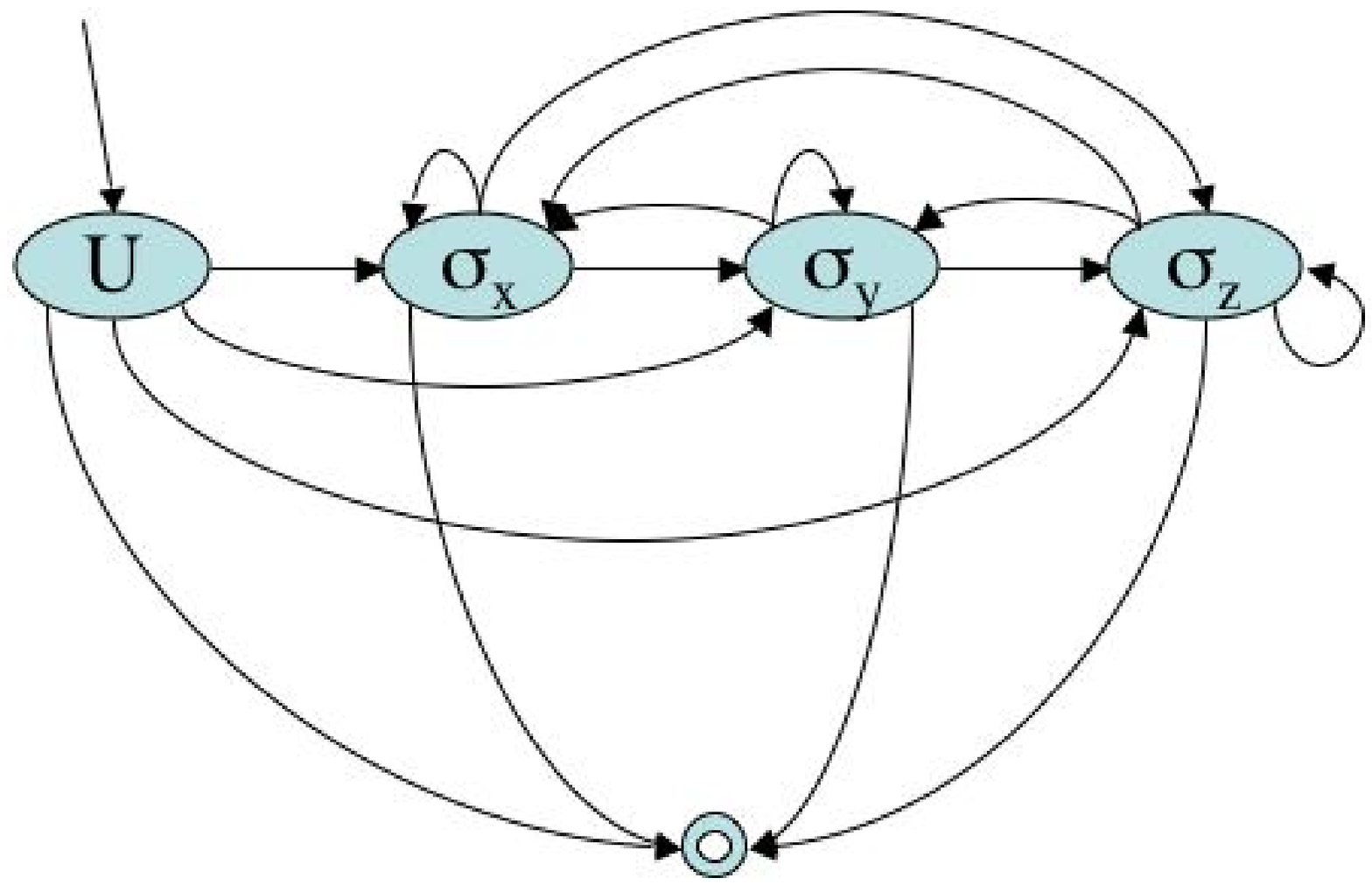}

\emph{\small Figure 9: Full simulation of $U$}
\end{center}

The number of states needed for the full simulation of $U$ is finite, hence $m$ has a finite number of states.
$\hfill \Box$

\end{document}